\documentclass[12pt,showpacs,preprintnumbers,superscriptaddress,nofootinbib]{revtex4}
\usepackage{multirow}
\usepackage{amssymb}
\usepackage{amsmath,graphicx}
\usepackage{dcolumn}
\usepackage{bm}
\usepackage{enumerate}
\usepackage{slashed}
\usepackage{epstopdf}
\usepackage[unicode]{hyperref}
\usepackage{braket}
\usepackage[usenames]{color}

\newcommand{\be}{\begin{equation}}
\newcommand{\ee}{\end{equation}}
\newcommand{\bea}{\begin{eqnarray}}
\newcommand{\eea}{\end{eqnarray}}
\newcommand{\ben}{\begin{enumerate}}
\newcommand{\een}{\end{enumerate}}
\newcommand{\bde}{\begin{widetext}}
\newcommand{\ede}{\end{widetext}}
\newcommand{\nn}{\nonumber}
\newcommand{\crn}{\nonumber \\}

\newcommand{\eq}{\eqref}
\newcommand{\al}{\alpha}
\newcommand{\la}{\lambda}

\newcommand{\om}{\omega}
\newcommand{\pa}{\partial}

\newcommand{\fr}{\frac}

\newcommand{\bc}{\begin{center}}
\newcommand{\ec}{\end{center}}
\newcommand{\Ga}{\Gamma}
\newcommand{\de}{\delta}
\newcommand{\De}{\Delta}
\newcommand{\ep}{\epsilon}

\newcommand{\La}{\Lambda}
\newcommand{\si}{\sigma}
\newcommand{\Si}{\Sigma}

\newcommand{\Om}{\Omega}
\begin{document}

\title{\boldmath Sphaleron in the first-order electroweak phase transition with the dimension-six Higgs operator}
\author{Vo Quoc Phong}
\email{vqphong@hcmus.edu.vn}
\affiliation{Department of Theoretical Physics, University of Science, Ho Chi Minh City 700000, Vietnam}
\affiliation{Vietnam National University, Ho Chi Minh City 700000, Vietnam}
\author{Phan Hong Khiem}
\email{phkhiem@hcmus.edu.vn}
\affiliation{Department of Theoretical Physics, University of Science, Ho Chi Minh City 700000, Vietnam}
\affiliation{Vietnam National University, Ho Chi Minh City 700000, Vietnam}

\author{Ngo Phuc Duc Loc}
\email{locngo148@gmail.com}
\affiliation{Department of Theoretical Physics, University of Science, Ho Chi Minh City 700000, Vietnam}
\affiliation{Vietnam National University, Ho Chi Minh City 700000, Vietnam}
\author{Hoang Ngoc Long}
\email{hoangngoclong@tdtu.edu.vn}
 \affiliation{Theoretical Particle Physics and Cosmology Research Group, Advanced Institute for Materials Science,
 Ton Duc Thang University, Ho Chi Minh City 700000, Vietnam}
 \affiliation{Faculty of Applied Sciences,
Ton Duc Thang University, Ho Chi Minh City 700000, Vietnam}
\date{\today }

\begin{abstract}
	By adding the dimension-six operator for the Higgs potential (denoted $\mathcal{O}_6$) in Standard Model, we have a first-order electroweak phase transition (EWPT) whose strength is larger than unity. The cutoff parameter  of the dimension-six Higgs operator ($\La$) is found to be in the range 593-860 GeV with the Wilson parameter equals to unity; it is also shown that the greater the $\La$, the lower the phase transition strength and the larger the Wilson parameter, the wider the domain of $\La $. At zero temperature, the sphaleron energy is calculated with a smooth ansatz and an ansatz with scale-free parameters, thereby we find that smooth profiles are not more accurate than profiles with scale-free parameters. Then, using the one-loop effective Higgs potential with the inclusion of $\mathcal{O}_6$ instead of all possible dimension-six operators, we directly calculate the electroweak sphaleron energy at finite temperature with the scale-free parameters ansatz and show that the decoupling condition is satisfied during the phase transition. Moreover, we can reevaluate the upper bound of the cutoff scale inferred from the first-order phase transition.  In addition, with the upper bound of the cutoff parameter (about 800-860 GeV), EWPT is a solution to the energy scale of the dimension-six operators. There is an extended conclusion that EWPT can only be solved at a large energy scale than that of SM.
	\end{abstract}
	\pacs{11.15.Ex, 12.60.Fr, 98.80.Cq}
	\maketitle
	Keywords:  Spontaneous breaking of gauge symmetries,
	Extensions of electroweak Higgs sector, Particle-theory models (Early Universe)	
	
%\tableofcontents

\section{INTRODUCTION}\label{secInt}

The Standard Model of particle physics (SM) has established many good results that agree with experiments and gave us a clear framework of how matter interacts with each other. The cornerstone of SM is the concept of symmetries such as Lorentz symmetry and gauge symmetry. We want the theory to be consistent with special relativity, which is the theory of space and time in the absence of gravitation, so we need a kind of external symmetry called Lorentz symmetry that transforms the coordinates of space and time. Motivated from this, we also want to have a kind of internal symmetry called gauge symmetry that transforms the dynamical fields themselves. The interesting thing is that gauge symmetry inevitably leads  to the existence of gauge bosons which mediate the interactions between matter (fermions). Furthermore, the Higgs mechanism of SM, which is basically changes of variables, can generate masses spontaneously without introducing by hands the mass terms of gauge bosons that might spoil the renormalizability of the theory. Starting from a few simple ideas, we got exactly what we want: A renormalizable theory that can describe matter and interactions (except gravity) in the flat background spacetime.

However, there are some fundamental problems that have not been solved by SM and one of them is the matter-antimatter asymmetry puzzle.  If we want a dynamical explanation for this rather than a conjecture of initial conditions, we have to investigate  electroweak phase transition (EWPT) process that happens in the early universe. In 1967, Sakharov proposed three conditions that a theory must have in order to solve the baryogenesis problem, which are baryon number violation, C and CP violation, and out-of-equilibrium condition. Nevertheless, EWPT within the context of SM does not offer good solutions since it does not satisfy sufficiently the last two conditions of  Sakharov. Therefore, theories beyond SM have to be taken into account. A common choice is considering that SM is only an effective theory valid up to a certain scale of energy called the cut-off scale (or the new-physics scale); this class of theory is called SMEFT (Standard Model Effective Field Theory). Of course, we expect that the results in SMEFT should reduce to those in SM in the limit that the cut-off scale goes to infinity, which is equivalent to the assumption that SM is valid at arbitrary high energy. Though, some care is required in making this statement since the Planck scale is believed to be the scale of energy where quantum gravity effects come into play.

SMEFT includes higher-dimensional, nonrenormalizable  operators such as dimension-five or dimension-six operators with a number of the so-called Wilson coefficients. There is nothing to prevent us from adding higher-dimensional operators, but there are also no solid physical reasons for adding them. The inclusion of these operators is just a phenomenological approach to achieve desired results; it is not originated from, for example, the defects in the theoretical framework of SM or a realization of a more reliable theory.

Now, we will focus on the aspects of SMEFT related to EWPT; and for this aim, the  dimension-six operators are simplest, good choices since, as we will see, they will affect some properties of EWPT in the way that we want. There are totally 20 possible dimension-six operators in the electroweak sector, but we want to consider only the dimension-six Higgs operator for the Higgs potential, denoted $\mathcal{O}_6$. There are some technical and physical reasons for this choice. Firstly, if we consider different dimension-six operators at a time, we will not be able to isolate the value of the cut-off scale $\La $ because there are many different Wilson coefficients. In other words, if we only consider $\mathcal{O}_6$, we can absorb the coefficient in front into the cut-off scale and hence calculations are simplified and definite statements about $\La $ could be made. Secondly, the $\mathcal{O}_6$ operator is the only dimension-six operator that can affect the form of the effective potential and can shift the strength of EWPT, which is connected to the third condition of Sakharov. Unfortunately, this operator does not have contributions to C and CP violations and therefore, we say that in the context of SMEFT the inclusion of $\mathcal{O}_6$ is required but not satisfactory to completely solve the baryogenesis problem. However, as we will see, the cut-off scale $\La $ is strongly constrained in order to have sufficient phase transition strength. Due to the fact that no new-physics phenomena are detected at the TeV scale, SMEFT with dimension-six operators is strongly suppressed and is not a prominent candidate to tackle the baryogenesis problem. Further inclusions of dimension-six operators that involve C and CP violations will not help unless the Wilson coefficient of $\mathcal{O}_6$ is unreasonably large compared to the values obtained from standard fitting methods \cite{1}.

In addition, since Higgs boson was discovered at the LHC, Particle Physics almost completed its mission that provide a more accurate understanding of mass. The EWPT problem will be one of the urgent issues. During the period from 1967 to the present, the EWPT has been calculated in SM as in Refs.~\cite{mkn,SME,michela} and in theories beyond SM or in many other contexts as in Refs.~\cite{2,BSM,majorana,thdm,ESMCO,elptdm,phonglongvan,SMS,dssm,munusm,lr,singlet,mssm1,twostep,1101.4665,jjg,Ahriche1,Ahriche2,Ahriche3,Fuyuto,span}. A familiar negative result is that the EWPT's strength is only larger than 1 at a few hundred GeV scale with the mass of Higgs boson must be less than $125$ GeV~\cite{mkn,SME,michela}. So far, the origin of a first-order EWPT can be heavy bosons or dark matter candidates ~\cite{2,majorana,thdm,ESMCO,elptdm,phonglongvan,epjc,zb,singlet,mssm1,twostep, chiang3}.

Another interesting finding is that EWPT is independent of the gauge. So the EWPT only needs to be calculated in the Landau gauge, which is sufficient and also physically suitable ~\cite{zb, 1101.4665, Arefe}. The damping effect in the thermal self-energy term or daisy loops have small distributions. The daisy loops are hard thermal loops, each of which contributes a factor of  $g^2T^2/m^2$~\cite{1101.4665} ($g$ is the coupling constant of $SU(2)$, $m$ is mass of boson), $m\sim 100$ GeV, $T\sim 100$ GeV, $g\sim 10^{-1}$ so $g^2T^2/m^2 \sim 10^{-2}$. When considering daisy loops, the improvement of effective potential leads to a reduction of the strength of EWPT roughly by a factor 2/3 ~\cite{r23}. Therefore, this distribution does not make a big change to the strength of EWPT or, in other words, it  is not the nature of EWPT.

There are some recent notable results with the $\mathcal{O}_6$ operator in the EWPT and sphaleron problems. The prediction about a first-order phase transition when there was an inclusion of $\mathcal{O}_6$ with different values of Higgs mass and cut-off parameters was established  in Ref.~\cite{13}. Refs.~\cite{chr,cde} also made a prediction of the first-order phase transition and estimated sphaleron energy at zero temperature, although the contribution of particles has not been calculated in detail yet. The tree-level EWPT sphaleron energy was solved by numerical methods at zero temperature with the dimension-six operators in Ref.~\cite{11}. Accordingly, in this paper, we attempt to constrain the range of the cut-off scale. While the lower bound the cut-off scale is determined by some mathematical arguments of the tree-level Higgs potential at zero temperature, the upper bound of the cut-off scale is determined from the requirement of EWPT's strength and is reassessed with the sphaleron energy at finite temperature using the scale-free parameters ansatz. In addition, with operators $\mathcal{O}_6$ associated with the scenario of cosmic inflation, Kusenko came up with a chart to calculate the baryon asymmetry of the Universe (BAU) at about $10^{-10}$ \cite{kusenko}.

There is one point we should clarify: the concept of a first-order EWPT should not be thought of as equivalent to the concept of departure from thermal equilibrium, although a first-order EWPT may indeed lead to the departure from thermal equilibrium. A first-order phase transition is defined as a kind of phase transition that happens when the two minima are separated by a potential barrier; this kind of phase transition is violent and the symmetry is broken via bubble nucleation. The departure from thermal equilibrium is defined more rigorously in the context of topological phase transition as $\Ga_{sph}\sim O(T,...) exp\left(\fr{- E_{sph}}{T}\right) \ll H_{rad}$ , where $\Ga_{sph}$ and $E_{sph}$ are the sphaleron rate and the sphaleron energy respectively, $H_{rad}$ is the Hubble expansion rate in the radiation-dominated era, and O(T,...) is a pre-exponential factor which is very hard to calculate. In the SM, this condition is translated into $v_c/T_c>1$ by using the approximate scaling relation $E_{sph}(T)\approx [v(T)/v] E_{sph}(T=0)$; however, in beyond SM models this scaling relation may break down. The interesting point of this paper is to investigate both the EWPT and the sphaleron solutions simultaneously so that we can understand the third condition of Sakharov better and, as we will see, we can assess the upper bound of the cut-off scale more thoroughly.

We have reviewed the current core results. Based on that, we will calculate EWPT and sphaleron in SM with $\mathcal{O}_6$ operator in this article, which is organized as follows. In section \ref{sec2}, we summarize some main features of SMEFT. In section \ref{sec3}, we derive the effective potential which has the contribution from the $\mathcal{O}_6$ operator. We analyze in detail the phase transition and find that the phase transition is first-order and show constraints on the cut-off scale. In section \ref{sec4}, electroweak sphalerons are calculated at zero and high temperature using the effective potential with the $\mathcal{O}_6$ operator. Finally, we summarize and make outlooks in section \ref{sec5}.

\section{SMEFT}\label{sec2}

The main idea of SMEFT is that SM is just an effective theory at low energy. In other words, the SM is essentially a leading order approximation of a more fundamental theory in the expansion of EFT with Lagrangian as follows

\be
\begin{aligned}
\mathcal{L}_{SMEFT}&=\mathcal{L}_{SM}+\fr{1}{\La }\sum_kC_k^{(5)}Q_k^{(5)}+\fr{1}{\La ^2}\sum_kC_k^{(6)}Q_k^{(6)}+\mathcal{O}\left(\fr{1}{\La ^3}\right),
\end{aligned}
\ee
where $\mathcal{L}_{SM}$ is the Lagrangian of SM, $\La $ is the cut-off scale (or new physics scale), the expansion factors $C_k$ are Wilson parameters and $Q_k$(s) are higher dimensional operators. Because the couplings have negative energy dimensions for operators of dimension-five or more, SMEFT is not renormalizable.

There are three conditions that SMEFT must satisfy \cite{12}: Firstly, the higher-order operators must satisfy the gauge symmetries in SM, which is $ SU (3) _C \times SU(2)_L \times U(1)_Y$. Secondly, the higher-order operators must contain all degrees of freedom of the  SM, either basic or composite. Thirdly, at low energy scales SMEFT must return to SM provided that there are no weak interacting light particles such as axions or sterlie neutrinos. According to Refs. \cite{hagi, 11}, we have totally 20 dimension-six operators that satisfy these requirements and their contributions at one-loop level is significant. However, in this paper we are interested in the $\mathcal{O}_6$ operator of the Higgs potential because it has an important effect to the EWPT process.

\section{Electroweak phase transition in SM with $\mathcal{O}_6$ operator}\label{sec3}

\subsection{Summary of calculating effective Higss potential}

The effective potential for quantum field theory was first introduced by Euler, Heisenberg and Schwinger \cite{3}. It was then applied to the spontaneous symmetry breaking survey by Goldstone, Salam, Weinberg and Jona-Lasinio \cite{4}. The 1-loop effective potential was calculated by Coleman and E.Weinberg in 1973 \cite {5}. For the case of multiple loops, one can refer to the calculations of Jackiw in 1974 \cite{6} and Iliopoulos, Itzykson and Martin in 1975 \cite{7}. In this article, we will stop at 1-loop level.

The one-loop Higgs effective potential when considering the contribution of heavy particles are usually derived in two ways. The first one is the functional approach, the second one is the perturbation approach. The high temperature potential is usually derived by using the Bose-Einstein or Fermi-Dirac distributions or using the finite-temperature Green function.

\subsubsection{The functional approach}

The first method was used by Coleman and Weinberg in 1973 \cite {5}, when using functional integrals and the finite temperature Matsubara Green function to derive the one-loop effective potential which has the contribution of all particles. The loops are fermion, gauge and neutral bosons loops and the external lines of these loops are the Higgs fields.

The action is given by
\be
S[\phi]=\int d^4x \mathcal{L}\{\phi(x)\}.
\ee

The generating functional $Z[j]$ is the transition amplitude from the vacuum in the far past to the one in the far future, with the source $j$,
\be
Z[j]=\braket{0_{out}|{0_{in}}}_j\equiv\int\mathcal{D}\phi\, \textrm{exp}\{i(S[\phi]+\phi j)\},
\ee
where
\be\label{notation}
\phi j\equiv \int d^4x \phi(x)j(x).
\ee
The connected generating functional $W[j]$ is defined by
\be
Z[j]\equiv \textrm{exp}\{iW[j]\}.
\ee

The effective action $\Ga [\overline{\phi}]$ is the Legendre transform of $W[j]$ as follows::
\be\label{effective action}
\Ga [\overline{\phi}]=W[j]-\int d^4x\fr{\de  W[j]}{\de  j(x)}j(x),
\ee
with
\be\label{classical field}
\phi_c=\overline{\phi}(x)\equiv\fr{\de  W[j]}{\de  j(x)}.
\ee

From Eqs. \eq{effective action}, \eq{classical field} and \eq{notation}, it is easy to prove that
\be
\begin{aligned}
\fr{\de \Ga [\overline{\phi}]}{\de \overline{\phi}}
&=\fr{\de  W[j]}{\de  j}\fr{\de  j}{\de \overline{\phi}}-\int d^4x \left[ j(x)+\overline{\phi}(x)\fr{\de  j(x)}{\de \overline{\phi}}\right]\\
&=\overline{\phi}(x)\fr{\de  j}{\de \overline{\phi}}-j-\overline{\phi} \fr{\de  j}{\de \overline{\phi}}\\
&=-j.
\end{aligned}
\ee
The above equation shows that there is no external source ($j$), $\overline{\phi}$ is the vacuum.

The Taylor expansion of connected generating and generating functional for $j$, yields
\be
\begin{cases}
Z[j]=\sum_{n=0}^\infty\fr{i^n}{n!}\int d^4x_1...d^4x_nj(x_1)...j(x_n)G_{(n)}(x_1,...,x_n)\\
iW[j]=\sum_{n=0}^\infty\fr{i^n}{n!}\int d^4x_1...d^4x_nj(x_1)...j(x_n)G_{(n)}^c(x_1,...,x_n)
\end{cases}.
\ee
The Green's functions $G_{(n)}$ are the sum of all Feynman diagrams with $n$ external lines. The expansion coefficients of $iW[j]$ are connected Green's functions and $G_{(n)}^c$ is the sum of all Feynman diagrams associated with $n$ external lines. Similarly, we have a Taylor expansion of the effective action in the $\overline{\phi}(x)$ field as follows:
\be\label{effective action Taylor expansion}
\Ga [\overline{\phi}(x)]=\sum_{n=0}^\infty\fr{1}{n!}\int d^4x_1...d^4x_n\overline{\phi}(x_1)...\overline{\phi}(x_n)\Ga ^{(n)}(x_1,...,x_n).
\ee

 $\Ga ^{(n)}$ is the sum of all one-particle-irreducible (1PI) Feynman diagrams with $n$ external lines. The Fourier transform of the effective action in $\overline{\phi}(x)$ is
\be
\begin{cases}
\Ga ^{(n)}(x_1,...,x_n)=\int\prod_{i=1}^n\left[\fr{d^4p_i}{(2\pi)^4}exp\{ip_ix_i\}\right](2\pi)^4\de {(4)}(p_1+...+p_n)\Ga ^{(n)}(p_1,...,p_n),\\
\tilde{\phi}(p)=\int d^4xe^{-ip.x}\overline{\phi}(x).
\end{cases}
\ee
From the two Fourier transforms above, the effective action in \eq{effective action Taylor expansion} is rewritten as follows
\be
\begin{aligned}
\Ga [\overline{\phi}(x)]&=\sum_{n=0}^\infty\fr{1}{n!}\int\prod_{i=1}^n\left[\fr{d^4p_i}{(2\pi)^4}\tilde{\phi}(-p_i)\right](2\pi)^4\de {(4)}(p_1+...+p_n)\Ga ^{(n)}(p_1,...,p_n).
\end{aligned}
\ee
The Fourier transform of $\phi_c$ is
\be
\tilde{\phi}_c(p)=\int d^4xe^{-ip.x}\phi_c=\phi_c(2\pi)^4\de {(4)}(p).
\ee
The effective potential becomes,
\be
\begin{aligned}
\Ga [\phi_c]&=\sum_{n=0}^\infty\fr{1}{n!}\int\prod_{i=1}^n\left[\fr{d^4p_i}{(2\pi)^4}\tilde{\phi}_c(-p_i)\right](2\pi)^4\de {(4)}(p_1+...+p_n)\Ga ^{(n)}(p_1,...,p_n) \\
&=\sum_{n=0}^\infty\fr{1}{n!}\phi_c^n(2\pi)^4\de {(4)}(0)\Ga ^{(n)}(p_i=0)\\
&=\int\sum_{n=0}^\infty\fr{1}{n!}\phi_c^n\Ga ^{(n)}(p_i=0) d^4x.
\end{aligned}
\label{mar1}
\ee

Meanwhile, the definition of the effective potential is
\be
\Ga [\phi_c]=-\int d^4xV_{eff}(\phi_c).\label{mar2}
\ee
Comparing the above two equations \eq{mar1} and \eq{mar2}, we get
\be\label{effective potential T=0}
V_{eff}=-\sum_{n=0}^\infty\fr{1}{n!}\phi^n\Ga ^{(n)}.
\ee

The Feynman rules at zero and finite temperature are shown in the table \ref{Feynman rules} \cite{8}, which is used to calculate the Green functions $\Ga ^{n}$
 or the sum of all diagrams.

The existence of dimension-six operators will affect many scattering and decay processes happening in particle accelerators, so the nonzero Wilson coefficients inferred from fitting methods will be the indication of new physics. However, most of the Wilson coefficients in the electroweak sector except $c_6$ are tightly constrained around zero \cite{1}. The difficulty in constraining $c_6$ lies in the fact that the Higgs potential in the SM is conventionally assumed to be of the form $\sim\phi^2+\phi^4$, but the Higgs self-coupling $\la $ is poorly measured. Moreover, with our choice of the form of $\mathcal{O}_6$ it will not affect $\la $ anyway; in other words, new physics cannot be inferred from the measurement of $\la $ but has important effects on EWPT. So it is reasonable to work solely with $\mathcal{O}_6$ in this paper.

We also note that this 1-loop effective potential has already included some dimension-six operators other than $\mathcal{O}_6$. For example, the 1-loop gauge boson term corresponding to $\mathcal{O}_r\sim W^2\phi^4$ is included in the $\Ga ^{(n)}$ function with $n=4$ external lines of the Higgs field. There are some exceptions like, for example, $\mathcal{O}_H\sim (\pa _\mu\phi^2)^2$ which cannot be implemented in the calculation of the effective potential but they have contributions to the sphaleron energy; we will not consider them in this paper.

\begin{table}[h]
	\caption{The Feynman rules at zero and finite temperature, $\beta$ is the Matsubara factor\cite{8}}
	\centering
	\begin{tabular}{|m{8em}|m{12em}|m{17em}|}
		\hline\hline
		Propagator  & {\boldmath $T=0$}  &{\boldmath$T\neq 0$}\\
		\hline
		Boson & $\fr{i}{p^2-m^2+i\ep }; p^\mu=(p^0,\mbox{\boldmath{$p$}})$ & $\fr{i}{p^2-m^2+i\ep };p^\mu=(2ni\pi\beta^{-1},\mbox{\boldmath{$p$}})$\\
		\hline
		Fermion & $\fr{i}{\slashed{p}-m+i\ep };p^\mu=(p^0,\mbox{\boldmath{$p$}})$
		& $\fr{i}{\slashed{p}-m+i\ep };p^\mu=((2n+1)i\pi\beta^{-1},\mbox{\boldmath{$p$}})$\\
		\hline
		Loop integral & $\int\fr{d^4p}{(2\pi)^4}$ & $\fr{i}{\beta}\sum_{n=-\infty}^{n=\infty}\int\fr{d^3p}{(2\pi)^3}$\\
		\hline
		Vertex function & $(2\pi)^4\de {(4)}\left(\sum_i p_i\right)$ & $-i\beta(2\pi)^3\de _{\Si \om _i}\de {(3)}\left(\sum_i\mbox{\boldmath{$p_i$}}\right)$\\
		\hline\hline
	\end{tabular}
	\label{Feynman rules}
\end{table}

The next work is from the Lagrangian of each field, we will calculate all their contributions by summing all the Green functions in Eq. \eq{effective potential T=0} and renormalization. We will have two components, one is quantum contributions, the other is thermal contributions. The thermal contribution is obtained by the finite temperature Matsubara Green function (or temperature propagators in the table \ref{Feynman rules}).

\subsubsection{The perturbation approach}

Firstly, we consider a toy model describing a self-interacting real scalar field and we see that this scalar field can be Higgs field. The scalar field satisfies the following equation of motion \cite{muka}
\be
\chi^{;\al }_{;\al }+V'(\chi)=0,\label{po}
\ee
where $V'=\fr{\pa V}{\pa \chi}$, potential $V$ is unknown. The field $\chi$ can always be decomposed into homogeneous and inhomogeneous components by adding a contribution from thermal fluctuation $\phi$:
\[
\chi(x)=\bar{\chi}(t)+\phi(x).
\]
Using Taylor series for $V(\bar{\chi}+\phi)$, we have
\[
V(\chi)=V(\bar{\chi})+\phi\fr{\pa  V(\bar{\chi})}{\pa \chi}+\fr{\phi^2}{2}\fr{\pa ^2 V(\bar{\chi})}{\pa ^2\chi}+\fr{\phi^3}{6}\fr{\pa ^3 V(\bar{\chi})}{\pa ^3\chi}+\mathcal{O}\left(\phi^4\right).
\]
Noting that $\fr{\pa  \phi}{\pa  \chi}=1$ and $\fr{\pa  \left(\fr{\pa  V(\bar{\chi})}{\pa \chi}\right)}{\pa  \chi}=0$, we get
\bea
\begin{split}
	V'(\chi)=V'(\bar{\chi})+\phi V''(\bar{\chi})+\fr{\phi^2}{2}V'''(\bar{\chi}).\label{ex}
\end{split}
\eea

Substituting (\ref{ex}) into (\ref{po}), and we only consider the potential component and averaging over space, we obtain
\be
\chi^{;\al }_{;\al }+V'(\bar{\chi})+\fr{1}{2}V'''(\bar{\chi})\left\langle \phi^2 \right\rangle=0.\label{exx}
\ee

The last two components in (\ref{exx}) can be rewritten as the derivative of an effective potential. In scalar field theory, we can get
\bea
\left\langle \phi^2 \right\rangle=\fr{1}{2\pi^2}\int \fr{k^2}{\sqrt{k^2+m^2_{\phi}(\bar{\chi})}}\left(\fr{1}{2}+n_{k}\right)dk.
\eea
Taking into account that $m^2_{\phi}=V''$. We can rewrite the third term in (\ref{exx}) as
\[
\begin{split}
\fr{1}{2}V'''(\bar{\chi})\left\langle \phi^2 \right\rangle&=\fr{1}{8\pi^2}\fr{\pa  m^2_{\phi}(\bar{\chi})}{\pa \bar{\chi}}\int \fr{k^2dk}{\sqrt{m^2_{\phi}(\bar{\chi})+k^2}}+\fr{1}{4\pi^2}\fr{\pa  m^2_{\phi}(\bar{\chi})}{\pa \bar{\chi}}\int \fr{k^2n_kdk}{\sqrt{m^2_{\phi}(\bar{\chi})+k^2}}\\
&=\fr{\pa  V^1_{\phi}}{\pa  \bar{\chi}}+\fr{\pa  V^2_{\phi}}{\pa  \bar{\chi}},
\end{split}
\]
where
\bea
V^1_{\phi}=\fr{1}{4\pi^2}\int k^2\sqrt{m^2_{\phi}(\bar{\chi})+k^2}dk=\fr{k \sqrt{k^2+m^2} \left(2 k^2+m^2\right)-m^4 \log\left[2 \left(k+\sqrt{k^2+m^2}\right)\right]}{32\pi^2},\crn
\label{v1}
\eea
\be
\fr{\pa  V^2_{\phi}}{\pa \bar{\chi}}=\fr{1}{4\pi^2}\fr{\pa m^2_{\phi}(\bar{\chi})}{\pa\bar{\chi}}\int \fr{k^2n_kdk}{\sqrt{m^2_{\phi}(\bar{\chi})+k^2}}.
\ee

Using Taylor expansion for (\ref{v1}), we obtain
\bea
\begin{split}
	V_{\phi}&=\fr{M^4}{16\pi^2}+\fr{m^2M^2}{16\pi^2}+\fr{m^4}{128\pi^2}+\fr{m^4}{32\pi^2}\ln m+\fr{m^4}{32\pi^2}\ln \fr{1}{2M}+\cdots\\
	&=\fr{m^4}{64\pi^2}\ln \left(\fr{m^2}{\mu^2}\right)+V_{\infty},
\end{split}
\eea
in which
\bea
V_{\infty}=\fr{M^4}{16\pi^2}+\fr{m^2M^2}{16\pi^2}-\fr{m^4}{32\pi^2}\ln \left(\fr{2M}{e^{1/4}\mu}\right)+\cdots
\eea
The occupation numbers $n_k$ are given by the Bose-Einstein formula, $\om _k=\sqrt{k^2+m^2_{\phi}}$
\[
\begin{split}
	\left\langle \phi^2\right\rangle_T=\fr{1}{2\pi^2}\int^{\infty}_{0}\fr{k^2dk}{\om _k e^{\fr{\om _k}{T}}-1}=\fr{T^2}{2\pi^2}\int^{\infty}_{\fr{m_{\phi}}{T}}\fr{\sqrt{\fr{\om ^2_k}{T^2}-\fr{m^2_{\phi}}{T^2}}}{e^{\fr{\om _k}{T}}-1}d\left(\fr{\om _k}{T}\right).
\end{split}
\]

$J^{(1)_{\pm}}$ has the following form
\bea
J^{(\nu)}_{\mp}(\al ,\beta)=\int^{\infty}_{\al }\fr{(x^2-\al ^2)^{\nu/2}}{e^{x-\beta}\mp 1}dx+\int^{\infty}_{\al }\fr{(x^2-\al ^2)^{\nu/2}}{e^{x+\beta}\mp 1}dx.
\eea

Substituting $\nu=1$, we obtain an expansion of $J^{1}_{\mp}$
\bea
\begin{split}
	J^{(1)}_{\mp}(\al ,\beta=0)&=2\int^{\infty}_{\al }\fr{(x^2-\al ^2)^{1/2}}{e^{x}\mp 1}dx\\
	&=\begin{cases}
		&\fr{1}{3}\pi^2-\fr{1}{2}\beta^2-\pi\sqrt{\al ^2-\beta^2}-\fr{1}{2}\al ^2\left(\ln\left(\fr{\al }{4\pi}\right)+C-\fr{1}{2}\right)+\mathcal{O}(\al ^2)\, ,\\
		&\fr{1}{6}\pi^2+\fr{1}{2}\beta^2+\fr{1}{2}\al ^2\left(\ln\left(\fr{\al }{\pi}\right)+C-\fr{1}{2}\right)+\mathcal{O}(\al ^2).\label{je}
	\end{cases}
\end{split}
\eea
Therefore we obtain
\bea
\begin{split}
	\left\langle \phi^2\right\rangle_T&=\fr{T^2}{2\pi^2}\int^{\infty}_{\fr{m_{\phi}}{T}}\fr{\sqrt{\fr{\om ^2_k}{T^2}-\fr{m^2_{\phi}}{T^2}}}{e^{\fr{\om _k}{T}}-1}d\left(\fr{\om _k}{T}\right)\\
	&=\fr{T^2}{4\pi^2}J^{(1)}_{-}\left(\fr{m_{\phi}(\bar{\chi})}{T},0\right)
\end{split}
\eea
and
\bea
\begin{split}
	\fr{1}{2}V'''(\bar{\chi})\left\langle \phi^2 \right\rangle_T&=m_{\phi}\fr{\pa m_{\phi}}{\pa \bar{\chi}}\fr{T^2}{4\pi^2}J^{1}_{-}\\
	&=\fr{\pa V^T_{\phi}}{\pa \bar{\chi}},\label{c}
\end{split}
\eea
with
\bea
\begin{split}
	V^T_{\phi}=\fr{T^4}{4\pi^2}\int^{\fr{m_{\phi}}{T}}_0\al  J^{1}_{-}(\al ,0)d\al =\fr{T^4}{4\pi^2}F_{-}\left(\fr{m_{\phi}}{T}\right).\label{b}
\end{split}
\eea
In case $\al =\fr{m_{\phi}}{T}$, we have
\bea
\fr{\pa V^T_{\phi}}{\pa \bar{\chi}}=\fr{\pa V^T_{\phi}}{\pa \al }\fr{\pa \al }{\pa \bar{\chi}}=\fr{\pa V^T_{\phi}}{\pa \al }\fr{\pa m_{\phi}}{T\pa \bar{\chi}},\label{a}
\eea
Substituting Eq.(\ref{a}) into Eq.(\ref{b}) yields
\bea
\begin{split}
	\fr{\pa V^T_{\phi}}{\pa \bar{\chi}}&=\fr{\pa m_{\phi}}{\pa \bar{\chi}}\fr{T^3}{4\pi^2}\int^{\fr{m_{\phi}}{T}}_0\fr{\pa [\al  J^{1}_{-}(\al ,0)]}{\pa \al }d\al \\
	&=\fr{\pa m_{\phi}}{\pa \bar{\chi}}\fr{T^2}{4\pi^2}m_{\phi}J^{1}_{-}(\al ,0),\label{d}
\end{split}
\eea
So Eq.(\ref{d}) is equivalent to Eq.(\ref{c}).

$V_{\infty}$ can be absorbed by a redefinition of constant in the original potential. The potential $V$ is
\be
V(\bar{\chi})=\fr{\la _0}{4}\bar{\chi}^4+\fr{m^2_0}{2}\bar{\chi}^2+\La _0,\label{v}
\ee

The final result, which includes both quantum and thermal contributions, is
\bea
V_{eff}=V+\fr{m^4_{\phi}(\bar{\chi})}{64\pi^2}\ln\left(\fr{m^2_{\phi}(\bar{\chi})}{\mu^2}\right)+\fr{T^4}{4\pi^2}F_{-}\left(\fr{m_{\phi}}{T}\right).\label{veff}
\eea
At zero temperature the last term in (\ref{veff}) will vanish. Then  the effective potential reduces to
\bea
V^{T=0}_{eff}=V+\fr{m^4_{\phi}(\bar{\chi})}{64\pi^2}\ln\left(\fr{m^2_{\phi}(\bar{\chi})}{\mu^2}\right).
\eea

Therefore, if we calculate the contributions of other particles and perform the renormalization process, we get the full 1-loop effective potential.

In addition, we have another way when the gauge and ring loops are taken into account by using functional methods, Nielsen identities and the small $\hbar$ expand as in articles \cite{twostep, 1101.4665}. However, we also see that the gauge and ring loops do not play a major role in the phase transition problem.

Ref.\cite{hagi} has shown the contributions of the dimension-six operators in the energy of sphaleron at one-loop level. In the calculation of the one-loop effective Higgs potential as mentioned in the functional approach, we have summed up all the contributions of one-loop diagrams. So in the effective potential, there are always the contributions of some dimension-six operators. We did not consider daisy loops, but we know that the thermal contributions of daisy loops are quite small, they can be ignored, this was also mentioned in the introduction section. Finally, when calculating the effective potential to investigate the EWPT and the sphaleron, we only need to use the $\mathcal{O}_6$ operator, regardless of the other dimension-six operators.

\subsection{Effective potential at $T=0$}
The  $\mathcal{O}_6$ operator is an addition to the Higgs potential, with all the contributions of SM bosons and top quark, Eq.\eq{effective potential T=0} becomes
\[
\begin{aligned}
V_{eff}(\phi,0)&=V_0(\phi,0)+V_1(\phi,0)+\mathcal{O}_{6}\\
&=\La _R-\fr{m_R^2}{2}\phi^2+\fr{\la _R}{4}\phi^4+\fr{1}{64\pi^2}\sum_{i=h,W,Z,t}n_im_i^4(\phi)ln\left(\fr{m_i^2(\phi)}{v^2}\right)+\fr{1}{8\La ^2}(\phi^2-v^2)^3.
\end{aligned}
\]

The $\mathcal{O}_6 $ operator was carefully chosen so that it does not affect the parameters and mass of the Higgs boson. Specifically,
\be
\begin{cases}
\mathcal{O}_{6}\Big|_{\phi=v}=0,\\
\fr{d\mathcal{O}_{6}}{d\phi}\Big|_{\phi=v}=\fr{6c_6}{8\La ^2}\phi(\phi^2-v^2)^2\Big|_{\phi=v}=0,\\
\fr{d^2\mathcal{O}_{6}}{d\phi^2}\Big|_{\phi=v}=\fr{6c_6}{8\La ^2}\left[(\phi^2-v^2)^2+4\phi^2(\phi^2-v^2)\right]\Big|_{\phi=v}=0.
\end{cases}
\ee

The normalization conditions for such the potential are:
\be
\begin{cases}
V_{eff}(v,0)=0, \\
V'_{eff}(v,0)=0,\\
V''_{eff}(v,0)=m_h^2.
\end{cases}
\ee
From these conditions, expanding the logarithmic functions, we get
\be\label{renormalization parameters}
\begin{cases}
\la _R=\fr{m_h^2}{2v^2}-\fr{1}{16\pi^2v^4}\sum_{i=h,W,Z,t}n_im_i^4\left[ln\left(\fr{m_i^2}{v^2}\right)+\fr{3}{2}\right],\\
m_R^2=\fr{m_h^2}{2}-\fr{1}{16\pi^2v^2}\sum_{i=h,W,Z,t}n_im_i^4,\\
\La _R=\fr{m_h^2v^2}{8}-\fr{1}{128\pi^2}\sum_{i=h,W,Z,t}n_im_i^4.
\end{cases}
\ee

\begin{figure}[h!]
	\centering
	\includegraphics[scale=0.8]{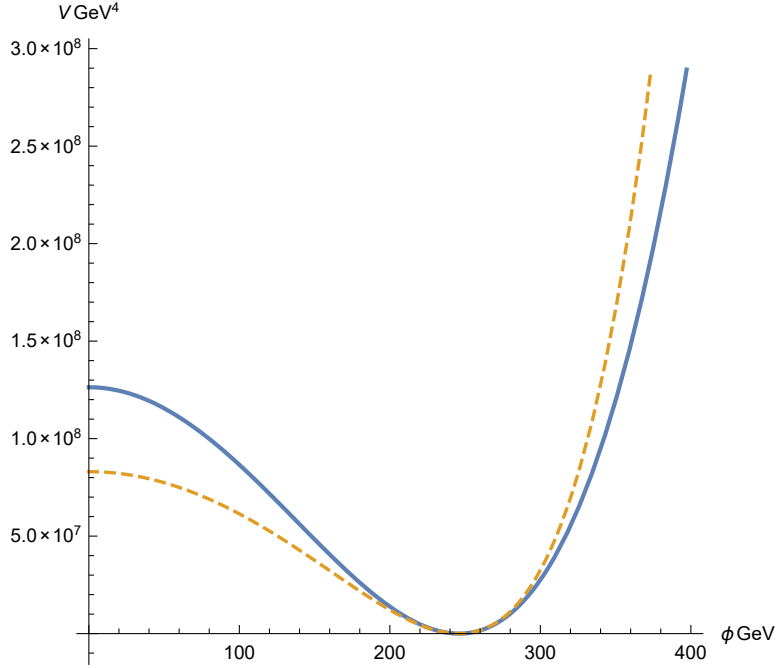}
	\caption{The vacuum contribution to effective potential. Dark line in blue: SM. Dashed line in orange: SM+ $\mathcal{O}_6$.}
	\label{vacuum contribution}
\end{figure}

The vacuum contribution of the effective potential is depicted in the figure \ref{vacuum contribution} with $\La =800$ GeV. The vacuum contribution ($T=0$) of SMEFT increases more rapidly and is larger than that of the SM in the range $\phi>246$ GeV. In contrast, the behavior of the potentials in the range $\phi<246$ GeV is opposite to that in the range $\phi>246$ GeV. This suggests that the minimum of SMEFT is more stable than that of the SM.

\subsection{Effective potential at $T\neq 0$}
The effective potential of SM when adding the $\mathcal{O} _6$ operator and expanding Eq.\eq{effective potential T=0} (or Eq.\eq{veff})  in detail is
\be\label{V_eff O6}
\begin{aligned}
V_{eff}(\phi,T)&=V_{eff}(\phi,0)+V_1(\phi,T)+\mathcal{O}_{6}\\
&=\fr{\la (T)}{4}\phi^4-ET\phi^3+D(T^2-T_0^2)\phi^2+\La (T)+\fr{1}{8\La ^2}(\phi^2-v^2)^3,
\end{aligned}
\ee
where
\bea
\la (T) &\equiv&\la _R+\fr{1}{16\pi^2v^4}\sum_{i=h,W,Z}n_im_i^4ln\left(\fr{a_iT^2}{v^2}\right),\crn
E&\equiv&\fr{1}{12\pi v^3}\sum_{i=h,W,Z}n_im_i^3=\fr{m_h^3+6m_W^3+3m_Z^3}{12\pi v^3},\crn
D&\equiv&\sum_{i=h,W,Z}\fr{n_im_i^2}{24v^2}-\fr{n_tm_t^2}{48v^2}=\fr{m_h^2+6m_W^2+3m_Z^2+6m_t^2}{24v^2},
\crn
T_0^2&\equiv &\fr{m_R^2}{2D},
\crn
\La (T)&\equiv&\La _R-\sum_{i=h,W,Z}\fr{n_i\pi^2T^4}{90}+n_t\fr{7\pi^2T^4}{720}=\La _R-\fr{41\pi^2T^4}{180}.
\nn
\eea

Using Eq. \eq {renormalization parameters}, we obtain  the explicit expressions of  the normalization parameters which was derived from the 1-loop effective potential at zero temperature, as follows
\be
\begin{aligned}
\la (T)&=\fr{m_h^2}{2v^2}-\fr{1}{16\pi^2v^4}\sum_{i=h,W,Z,t}n_im_i^4\left[ln\left(\fr{m_i^2}{v^2}\right)+\fr{3}{2}\right]
+\fr{1}{16\pi^2v^4}\sum_{i=h,W,Z}n_im_i^4ln\left(\fr{a_iT^2}{v^2}\right)\\
&=\fr{m_h^2}{2v^2}-\fr{1}{16\pi^2v^4}\sum_{i=h,W,Z,t}n_im_i^4ln\left(\fr{m_i^2}{A_iT^2}\right),
\end{aligned}
\nn
\ee
\bea
T_0^2 &=&\fr{1}{2D}\left[\fr{m_h^2}{2}-\fr{1}{16\pi^2v^2}\sum_{i=h,W,Z,t}n_im_i^4\right]=
\fr{1}{D}\left[\fr{m_h^2}{4}-\fr{1}{32\pi^2v^2}\sum_{i=h,W,Z,t}n_im_i^4\right],\crn
\La (T)&=&\fr{m_h^2v^2}{8}-\fr{1}{128\pi^2}\sum_{i=h,W,Z,t}n_im_i^4-\fr{41\pi^2T^4}{180}.
\nn
\eea

Here, we have put
\be
\begin{cases}
\text{Boson: }ln(A_B)\equiv ln(a_b)-3/2=3.9076, \\
\text{Fermion: }ln(A_F)\equiv ln(a_f)-3/2=1.1351.
\end{cases}
\ee

The transition-ending temperature is the temperature at which the initial minimum becomes a maximum:
\[
T_0'^2=T_0^2-\fr{3v^4}{8D\La ^2}.
\]

The above expansion for a high-temperature effective potential will be better than $5\%$ if $\fr{m_{boson}}{T}< 2.2$ ~\cite{5percent}, where $m_{boson}$
 is the relevant boson mass. The mass range of SM particle in the EWPT calculations always fit with this because the critical temperature is about 100 GeV.

\begin{figure}[h!]
	\centering
	\includegraphics[scale=0.9]{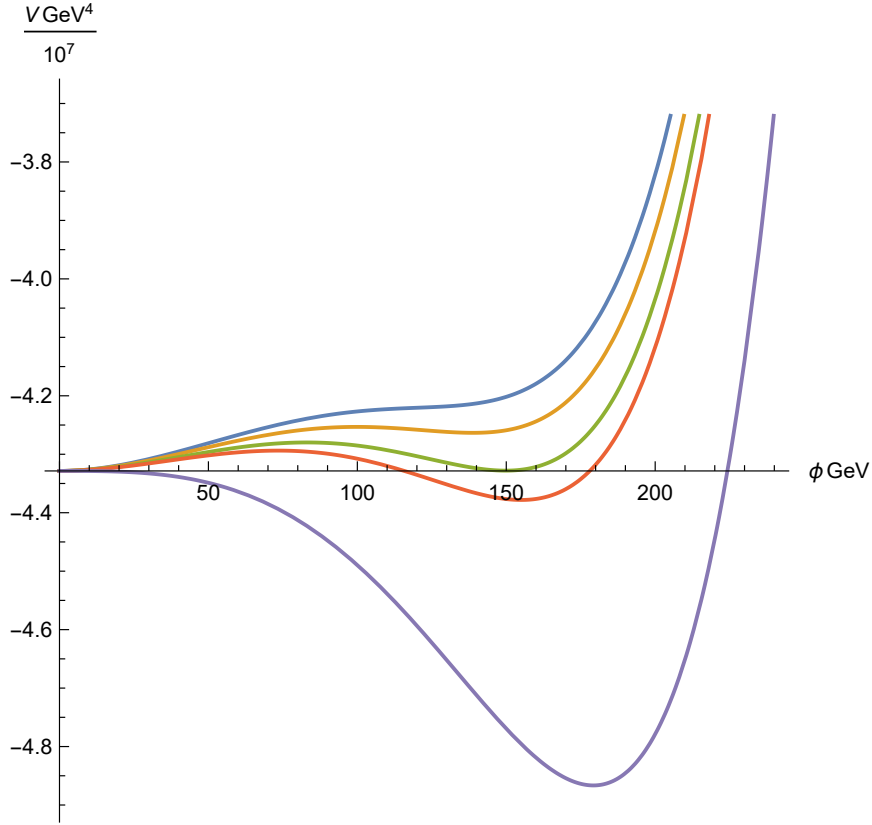}
	\caption{The effective potential at finite temperature in SM with $\mathcal{O}_6$ ($\La =800$ GeV). Blue line: $T=120$ GeV. Orange line: $T=T_1=119.35$ GeV. Green line: $T=T_c=118.56$ GeV. Red line: $T=118$ GeV. Purple line: $T=T_0'=113.36$ GeV (the end of phase transition).}
	\label{vacuum and thermal contributions in SM with O6}
\end{figure}

The effective potential is drawn in the figure \ref{vacuum and thermal contributions in SM with O6}. It appears the non-zero second minimum and the value of this minimum grows as temperature decreases. Between two minima, a barrier appears which is a sign of first-order phase transition.

The results of phase transition in the range $700$ GeV $\leq\La \leq 1000$ GeV, are summarized in the table \ref{cdcp}. An interesting conclusion is that
when $\La $ increases, $T_1$ (the second minimum starts to appear from here) and the critical temperature ($ T_c $) approach each other. As expected, the bigger the cutoff scale, the smaller the phase transition strength. When $\La  \rightarrow \infty$, the transition strength returns to the value of the SM. The maximum possible value of the cut-off scale at which the phase transition strength is greater than unity is about 860 GeV.
\begin{table}[!ht]
		\caption{Phase transition strengths with different values of $\La $.}
	\bc
		\begin{tabular}{m{4em}m{5em}m{5em}m{5em}m{3em}}
			\hline\hline
			$\La $ (GeV) & $T_1$ (GeV) & $T_c$ (GeV)& $v_c$ (GeV) & $v_c/T_c$\\
			\hline
			1000 & 132.89 & 132.82 & 86.53 & 0.65\\
			\hline
			900 & 127.07 & 126.68 & 114.22 & 0.9\\
			\hline
			880 & 125.74 & 125.24 & 120.79 & 0.96\\
			\hline
			870 & 125.08 & 124.48 & 124.38 & 0.999\\
			\hline
			860 & 124.08 & 123.71 & 127.81 & 1.03\\
			\hline
			840 & 122.49 & 122.08 & 135.21 & 1.11\\
			\hline
			820 & 120.97 & 120.38 & 142.36 & 1.18\\
			\hline
			800 & 119.35 & 118.56 &	149.78 & 1.26\\
			\hline
			700 & 111.02 & 107.53 & 184.72 & 1.72\\
			\hline\hline
		\end{tabular}\label{cdcp}
	\ec
\end{table}

%%%%den day %%%%%
\subsection{The lower bound of $\La $}

From pure mathematical arguments, we can also find the lower bound of $\La $. The tree-level Higgs potential in SM with the inclusion of $\mathcal{O}_6$  is
\[
V^{SM+\mathcal{O}_6}_{tree}(\phi)=\Om -\fr{\mu^2}{2}\phi^2+\fr{\la }{4}\phi^4+\fr{c_{6}}{8\La ^2}(\phi^2-v^2)^3,
\]
where $\Om $ is a cosmological constant and $\la >0$. Here we have temporarily written down explicitly the Wilson coefficient of $\mathcal{O}_6$ since we will also have the chance to discuss its sign. We find the local extremes of potential by calculating its first derivative:
\be\label{first derivative of higgs-tree}
V'(\phi)=-\mu^2\phi+\la \phi^3+\fr{3c_{6}}{4\La ^2}\phi(\phi^2-v^2)^2=0.
\ee
\[
\Rightarrow \begin{cases}
\phi=0 \,,\\
\mathcal{A}\phi^4+(\la -2\mathcal{A}v^2)\phi^2-\mu^2+\mathcal{A}v^4=0
\end{cases},
\]
here we have set $\mathcal{A}\equiv \fr{3c_{6}}{4\La ^2}$ for brevity, and the discriminant $\De $ of the quartic equation is
\be\label{Delta}
\De =\la ^2-4\mathcal{A}\la  v^2+4\mathcal{A}\mu^2.
\ee
On the other hand, since $\mathcal{O}_6$ does not affect three normalized conditions: $V (v)=0, V'(v)= 0, V''(v)=m^2_h$ so the coefficients in the Higgs potential are still the same form as those of SM:
\be\label{coefficients of tree-level Higgs potential}
\begin{cases}
\mu^2=\la  v^2 \,,\\
m_h^2=2\la  v^2\,,\\
\Om =\fr{1}{4}\la  v^4
\end{cases}.
\ee
Substituting $\mu^2=\la  v^2$ into Eq.\eq{Delta}, we get
\[
\De =\la ^2>0.
\]
The solutions of the quartic equation are
\[
\phi^2=\fr{2\mathcal{A}v^2-\la \pm \la }{2\mathcal{A}}\rightarrow\begin{cases}
\phi^2=v^2\,,\\
\phi^2=\fr{\mathcal{A}v^2-\la }{\mathcal{A}}
\end{cases}.
\]
At zero temperature, we expect that Eq.\eq{first derivative of higgs-tree} should have three solutions corresponding to the three extremes at
$\phi=0$ and $\phi=\pm v$, so we must have the following condition:
\[
\fr{\mathcal{A}v^2-\la }{\mathcal{A}}<0.
\]
From Eq.\eq{coefficients of tree-level Higgs potential}, with  $\la =\fr{m_h^2}{2v^2}$, we get
\be
\begin{aligned}
\mathcal{A}&<\fr{\la }{v^2}\, ,\\
\Rightarrow \fr{3c_{6}}{4\La ^2}&<\fr{\la }{v^2}=\fr{m_h^2}{2v^4}\,,\\
\Rightarrow \La ^2&>\fr{3c_{6}v^4}{2m_h^2}\,,\\
\Rightarrow \La &>\sqrt{1.5c_{6}}\fr{v^2}{m_h}.
\end{aligned}
\label{mar3}
\ee

We see that the case $c_6<0$ must be removed. With $c_6=1$ and substituting the experimental values $v= 246$ GeV and $m_h=125$ GeV into \eq{mar3}, we get $\La > 593$ GeV. Therefore, the larger the Wilson parameter, the bigger the lower bound. Overall, we have:

\be
\sqrt{1.5c_{6}}\fr{v^2}{m_h}<\La <860.\sqrt{c_{6}} \Longleftrightarrow  593 \text{ GeV}<\fr{\La }{\sqrt{c_{6}}}<860 \text{ GeV}.\label{mien}
\ee

\begin{figure}[!ht]
	\centering
	\includegraphics[scale=1.2]{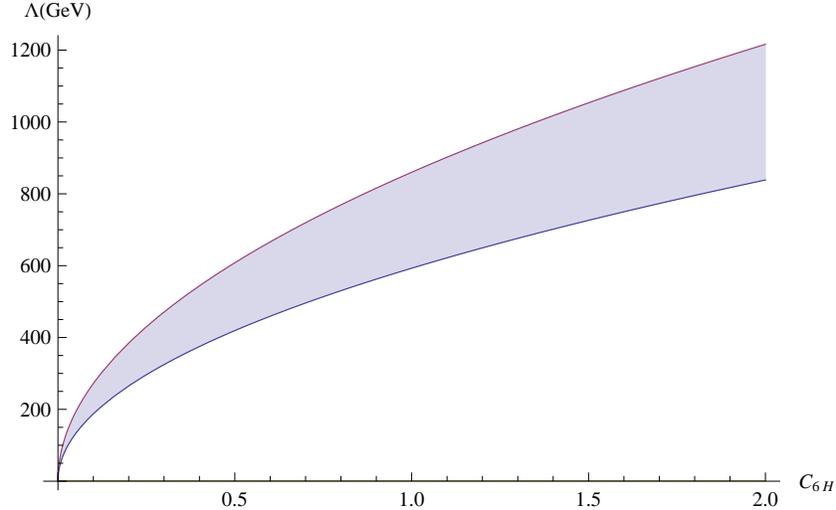}
	\caption{The shaded area is the allowed range for the cut-off parameter according to the Wilson parameter $c_{6}$.}
	\label{wilsonhinh}
\end{figure}

Thus, the larger the Wilson parameter value, the wider the domain of $\La $ in order to have the first-order phase transition, as shown in the figure \ref{wilsonhinh}. 

\section{The energy of sphaleron}\label{sec4}

In a non-Abelian gauge theory such as the electroweak theory there is another interesting kind of transition called topological phase transition, which is the transition of the field configurations between topologically distinct vacua. When such a transition occurs, the baryon number will be violated and it opens up the possibility for the electroweak theory to satisfy the first condition of Sakharov. At zero temperature, the instanton process is strongly suppressed due to the smallness of the weak coupling \cite{muka}. On the other hand, at high temperature the sphaleron process dominates and the transition rate is proportional to the Boltzmann factor as $\Ga_{sph}\sim O(T,...) \textrm{exp}(-E_{sph}/T)$. When the temperature is high, the "height" of the potential, which is the energy of sphaleron, is small and thermal equilibrium can be obtained. However, when the temperature is low, the height of the potential barrier will rise and the sphaleron rate must be strongly suppressed. The universe also expands while this phase transition occurs, so the condition for the departure from thermal equilibrium is $\Ga_{sph} \ll H_{rad}$.

According to Ref.~\cite {11}, the sphaleron energy at zero temperature at tree level is calculated by including all relevant dimension-six operators. However, the first difficulty is that it is  unclear whether the Wilson parameters are positive or negative; the second difficulty is that the equations of motion are temperature-dependent when considering EWPT and hence are even more difficult to be solved.

\subsection{Sphaleron ansatz}

Manton and Klinkhamer \cite{10} have found the ansatz approach to find an approximate sphaleron solution without fully solving all the  equations of motion. The contribution to the sphaleron energy of the hypercharge gauge field is found to be very small, so we will not consider them (i.e. $B_\mu=0$) \cite{10}. We also choose the temporal gauge, so $W^a_0=0$. Let $n_i \equiv \fr{x^i}{r}$ where $r$ is the radial coordinate in the spherical coordinates and assume that sphaleron has a spherically symmetric form. The sphaleron ansatzes then have the following forms \cite{11}
\be\label{ansatz}
\begin{cases}
H(r)=\fr{v}{\sqrt{2}}h(r)in_a\si^a\left(\begin{matrix}
0\\
1
\end{matrix}\right)\,,\\
W^a_i(r)=\fr{2}{g}\ep ^{aij}n_j\fr{f(r)}{r}\,,
\end{cases}
\ee
with the boundary conditions are
\be\label{BCs}
\begin{cases}
h(r\rightarrow 0)=f(r\rightarrow 0)=0,\\
h(r\rightarrow\infty)=f(r\rightarrow\infty)=1.
\end{cases}
\ee
The functions $f(r)$ and $h(r)$ are called radial functions (or profile functions). The first boundary condition is intended to avoid singularity at $r=0$, and the second boundary condition is to ensure that the ansatzes are asymptotic to the form of the fields at infinity\cite{10}. As we will see, in this way the equations of motion will be converted to the equations of $h(r)$ and $f(r)$ and everything will be much simpler.

Next, in order to calculate the sphaleron energy, we need to solve all the equations of motion to find $h(r)$ and $f(r)$ and then substitute these functions into the spahleron energy functional. However, solving these equations of motion is also very complicated and can only be done by numerical method  as in Ref. \cite{11}. To be more streamlined we can assume the above profiles by spherical functions, and refine these functions through free parameters. The calculation steps are shown in the next section.

\subsection{Contributions to the sphaleron energy}
The sphaleron energy functional takes the form
\be\label{e}
E_{sph}=\int d^3x\left[\fr{1}{4}W^a_{kl}W^{a}_{kl}+(D_kH)^\dagger(D_kH)+V_{Higgs}(H,T)\right].
\ee

\begin{itemize}
	\item \emph{The first term} in Eq.\eq{e} is the contribution of the isospin gauge fields $\fr{1}{4}W^a_{kl}W^{a,kl}$:\\
    Substituting the ansatzes in \eq{ansatz}  into the first term in \eq{e} yields
	\[
	\fr{1}{4}W^a_{kl}W^{a,kl}=\fr{2}{g^2r^6}[2f'^2r^4+4r^2f^2(1-f)^2].
	\]
	So the sphaleron energy contribution of the gauge fields is
	\be
	E_{sph}^{gauge}=\fr{4\pi}{g^2}\int_0^\infty dr\left[4f'^2+\fr{8f^2(1-f)^2}{r^2}\right].
\label{mar4}
	\ee
	We can use the following dimensionless distance variable
	\[
	\xi\equiv gvr\rightarrow d\xi =gvdr\rightarrow\fr{df}{dr}=\fr{df}{d\xi}\fr{d\xi}{dr}=gv\fr{df}{d\xi}.
	\]
	So we get the sphaleron energy of the gauge fields in terms of the variable $\xi$ as follows
	\be
	\begin{aligned}
	E_{sph}^{gauge}&=\fr{4\pi v}{g}\int_0^\infty d\xi\left[4\left(\fr{df}{d\xi}\right)^2+\fr{8f^2(1-f)^2}{\xi^2}\right].
	\end{aligned}
\label{mar5}
	\ee
	\item \emph{The second term} in Eq.\eq{e} is the kinematic of the Higgs field $(D_k H)^\dagger (D_kH)$.
	
	Following the same procedure as we did in the previous case, the contribution of the kinematic term of the Higgs field to the sphaleron energy is
	\be
	\begin{aligned}
	E_{sph}^{\text{kinematic Higgs}}
	&=\fr{4\pi v}{g}\int_0^\infty d\xi\left[\fr{\xi^2}{2}\left(\fr{dh}{d\xi}\right)^2+h^2(1-f)^2\right].
	\end{aligned}
\label{mar6}
	\ee
	\item \emph{The third term} in Eq.\eq{e} is the Higgs potential with $\mathcal{O}_6$ at tree-level.
	\be
	\begin{aligned}
	V_{Higgs}(H,T=0)&=-\mu^2(H^\dagger H)+\la (H^\dagger H)^2+\Om +\fr{1}{\La ^2}\left(H^\dagger H-\fr{v^2}{2}\right)^3\\
	&=-\fr{\mu^2v^2}{2}h^2+\fr{\la  v^4}{4}h^4+\Om +\fr{v^6}{8\La ^2}(h^2-1)^3\\
	&= \fr{\la  v^4}{4}(h^2-1)^2+\fr{v^6}{8\La ^2}(h^2-1)^3.
	\end{aligned}
	\ee
\label{mar7}
\end{itemize}

 It is emphasized that, according to  Ref. \cite{Ahriche1}, the contributions to  the sphaleron energy from  the first and the second terms in Eqs. \eq{mar5} and  \eq{mar6}, respectively, are  large. The difference between sphaleron energy at zero temperature and sphaleron energy at finite temperature lies solely in the Higgs potential. The form of contributions from kinematic terms remains the same, but the Higgs profile function $h(r)$ and the dimensionless distance must be rescaled when calculating sphaleron energy at nonzero temperature.

\subsection{Profile functions}

There are two kinds of ansatz: the ansatz with scale-free parameters firstly  introduced in \cite{10}, and the smooth ansatz motivated from kink-type solutions \cite{11,46}. We will use both of these two types to calculate the sphaleron energy at zero temperature and compare it to the values obtained from numerical calculations, then we will use the better ansatz to calculate sphaleron energy at finite temperature.

We use the following ansatz
\ben
\item Ansatz $a$: $f$ and $h$ with scale-free parameters
	\be\label{f}
	f^a(\xi)=\begin{cases}
	\fr{\xi^2}{2a^2}, \hspace{1cm} \xi\leq a \,,\\
	1-\fr{a^2}{2\xi^2},\hspace{1cm} \xi\geq a ;
	\end{cases}
	\ee	
	\be\label{h}
	h^a(\xi)=\begin{cases}
	\fr{4\xi}{5b},\hspace{1cm}\xi\leq b\,,\\
	1-\fr{b^4}{5\xi^4}\hspace{1cm}\xi\geq b\,.
	\end{cases}
	\ee
with $a,b$ are scale-free parameters.

\item Ansatz $b$: $f$ and $h$ are smooth functions
	\be\label{smooth profile}
	f^b(\xi)=tanh(A\xi)\hspace{1cm};\hspace{1cm}h^b(\xi)=tanh(B\xi),
	\ee
where $A$ and $B$ are free parameters.
\een

Note that we must select the profile functions that satisfy the boundary conditions in \eq{BCs}.

\subsection{Sphaleron energy at zero temperature at tree level with $\mathcal{O}_6$}

The zero temperature sphaleron energy at tree-level with $\mathcal{O}_6$ is
\be
\begin{aligned}
E_{sph,0}^{tree}=\fr{4\pi v}{g}\int_0^\infty d\xi &\Bigg[4\left(\fr{df}{d\xi}\right)^2+\fr{8}{\xi^2}f^2(1-f)^2+\fr{\xi^2}{2}\left(\fr{dh}{d\xi}\right)^2+h^2(1-f)^2\\
&+\fr{\xi^2}{g^2}\left(\fr{\la }{4}(h^2-1)^2+\fr{v^2}{8\La ^2}(h^2-1)^3\right)\Bigg].
\end{aligned}\label{91}
\ee
As mentioned in the introduction, the Wilson coefficient of $\mathcal{O}_6$ can be absorbed into $\La $. Assuming that the Higgs potential in the  SM has the standard form $\sim \phi^2 + \phi^4$, we have $\la  \approx 0.1$. Because choosing the $\mathcal{O}_6$ operator does not change the Higgs mass form, so we still have the value of $\la $ like in the SM.

\ben
\item Ansatz $a$ (Eqs. \eq{f} and \eq{h})

With ansatz $a$, Eq.\eq{91} becomes:
\be
E_{sph,0}^{tree}=\fr{832}{315 a}-\fr{104a^4}{1155b^3}+\fr{704a^3}{2625b^2}+\fr{4b}{43509375}\left[1657500+\fr{b^2}{g^2\La ^2}\left(447100\la \La ^2-137997v^2\right)\right].\label{92}
\ee
To minimize this energy according to the parameters $a$ and $b$, we get the energy value table of sphaleron energy in the cut-off scale as table \ref{freean}.

\begin{table}[!h]
	\caption{Sphaleron energy at zero temperature with $\mathcal {O}_6 $ in ansatz $a$, in units of $4\pi v/g \approx 4,738$ TeV.}
	\bc
		\begin{tabular}{m{5em}m{10em}m{3em}m{3em}}
			\hline\hline
			$\La $ [GeV] & $E_{sph,0}^{tree}[\times\fr{4\pi v}{g}TeV]$ & a & b\\
			\hline
			$\infty$ & 2.024 & 2.36 & 2.23\\
			\hline
			1000 & 2.002 & 2.39 & 2.33\\
			\hline
			900 & 1.997 & 2.41 & 2.36 \\
			\hline
			800 & 1.989 & 2.42 & 2.40\\
			\hline
			700 & 1.976 & 2.45 & 2.47\\
			\hline\hline
		\end{tabular}\label{freean}
	\ec
\end{table}

\item Ansatz $b$ (Eq.\eq{smooth profile})

With ansatz $b$, we get the energy value table of sphaleron energy in the cut-off scale as table \ref{smooth ansatz}.

\begin{table}[h]
	\caption{Sphaleron energy at zero temperature with $\mathcal {O}_6 $ in ansatz $b$, in units of $4\pi v/g \approx 4,738$ TeV.}
	\centering
	\begin{tabular}{m{5em}m{10em}m{3em}m{3em}}
		\hline
		$\La $ [GeV] & $E_{sph,0}^{tree}[\times\fr{4\pi v}{g}TeV]$ & A & B\\
		\hline
		$\infty$ & 2.144 & 0.22 & 0.48\\
		\hline
		1000 & 2.126 & 0.22 & 0.47\\
		\hline
		900 & 2.122 & 0.21 & 0.45\\
		\hline
		800 & 2.115 & 0.21 & 0.44\\
		\hline
		700 & 2.104 & 0.21 & 0.43\\
		\hline
	\end{tabular}
	\label{smooth ansatz}
\end{table}
\een

\begin{figure}[!h]
		\centering
		\includegraphics[scale=1]{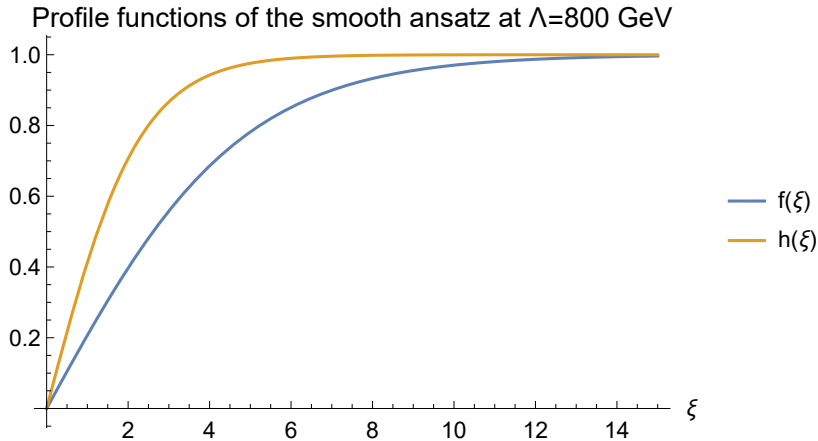}
		\caption{The tree-level profile functions for the Higgs field and the isospin gauge fields at zero temperature using ansatz $b$ with $\La =800$ GeV. Orange line: the Higgs field. Blue line: the isospin gauge fields.}
		\label{smooth}
	\end{figure}

The profile functions of ansatz $b$ are depicted in figure \ref{smooth}.
\bc
\ec

With ansatz a, the sphaleron energy in the SM ($\La =\infty$) is about $2.024 \times \fr{4 \pi v}{g}$ (TeV). Comparing this value to the value obtained from numerical method in Ref. \cite{11}, we find that ansatz $a$ is in good agreement and the difference is only about 5.6 \%, while ansatz $b$ gives a difference of about 11.9\%. In addition, the sphaleron energy which has the contribution of $\mathcal {O} _6 $ at zero temperature, with $\La $ ranging from $590$ GeV to $860$ GeV, is in the range $[1.8-2]\times\fr{4\pi v}{g}$ (TeV) \cite{11}. So according to the tables \ref{freean} and \ref{smooth ansatz}, ansatz $a$ also gives better results. Therefore, we will only use the ansatz $a$ to calculate the sphaleron energy at nonzero temperature at one-loop level.

\subsection{Sphaleron energy at finite temperature at 1-loop level with $\mathcal{O}_6$}

Substituting the 1-loop effective potential (Eq. \eq{V_eff O6}) into the  sphaleron energy functional, we obtain the sphaleron energy at the temperature $T$:
\[
E_{sph}^{1-loop}(T)=\fr{4\pi v}{g}\int_0^\infty d\xi \Bigg[4\left(\fr{df}{d\xi}\right)^2+\fr{8}{\xi^2}f^2(1-f)^2+\fr{\xi^2}{2}\left(\fr{dh}{d\xi}\right)^2+h^2(1-f)^2
+\fr{\xi^2}{g^2v^4}V_{eff}(h,T)\Bigg].
\]
Taking the variations with respect to $f$ and $h$, we get the following system of equations of motion
\be \label{EOM}
\begin{cases}
\xi^2\frac{d^2f}{d\xi^2}=2f(1-f)(1-2f)-\frac{\xi^2}{4}h^2(1-f)\\
\frac{d}{d\xi}\left(\xi^2\frac{dh}{d\xi}\right)=2h(1-f)^2+\frac{\xi^2}{g^2v^4}\frac{\partial V_{eff}(h,T)}{\partial h}
\end{cases}.
\ee

The profile functions do not necessarily satisfy the nonlinear equations Eq.(\ref{EOM}). However, as we shall see, finding an ansatz satisfying the equations of motion at a certain asymptotic limit, especially in the sphaleron core region, will give better results. The "ansatz a" satisfies these equations in the sphaleron core region ($\xi\rightarrow 0$) and this ansatz is also suitable to calculate sphaleron energy at finite temperature.

In the electroweak phase transition in the section \ref{sec3}, the effective potential depends on $\phi$.  In the sphaleron problem, we want to express it through the profile functions $h$ and $f$. From the ansatz of the Higgs doublet and gauge field in the equation \eq{ansatz}, we obtain
\[
H^\dagger H=\fr{v^2h^2}{2}.
\]
On the other hand, ignoring the contribution of the Goldstone bosons, the Higgs doublet is
\[
H=\fr{1}{\sqrt{2}}\left(\begin{matrix}
0\\
\phi
\end{matrix}\right)\rightarrow H^\dagger H=\fr{\phi^2}{2}.
\]
So we have the following relationship
\[
\phi=vh.
\]
The above equation allows us to convert the effective potential in $\phi$ to that in $h$.

When calculating the sphaleron energy at temperature $T$, to be comparatively consistent with the sphaleron energy at zero temperature, ansatz $a$ must be used. However, it should be noted that we must rescale the variable $\xi$ and the radial function $h(\xi)$ as follows:
\[
\tilde{\xi}=\fr{\xi}{s}; \hspace{1cm}
\tilde{h}=sh;\hspace{1cm} \text{with }s\equiv\fr{v(T)}{v}.
\]

To avoid cumbersome, we will ignore the tilde symbol and automatically perform the above operation. The effective potential is then given by
\[
\begin{aligned}
V_{eff}(h,T)&=\fr{\la (T)}{4}\phi^4-ET\phi^3+D(T^2-T_0^2)\phi^2+\La (T)+\fr{1}{8\La ^2}(\phi^2-v^2)^3\\
&= \fr{\la (T)v^4}{4}s^4h^4-ETv^3s^3h^3+D(T^2-T_0^2)v^2s^2h^2+\fr{v^6}{8\La ^2}(s^2h^2-1)^3+\La (T)\\
&=\fr{\la (T)v^4}{4}s^4(h^4-1)-ETv^3s^3(h^3-1)+D(T^2-T_0^2)v^2s^2(h^2-1)\\
&+\fr{v^6}{8\La ^2}\Big[(s^6(h^6-1)-3s^4(h^4-1)+3s^2(h^2-1)\Big]+C(T).
\end{aligned}
\]
In the above equation, in the first line, we only rewrite the effective potential in the section \ref{sec3}; in the second line, we substitute $\phi=vh$ and rescale the radial function $h$; in the third line, we add and remove the temperature-dependent constants and set it to $C(T)$. The sphaleron energy will be rewritten as
\[
\begin{aligned}
E_{sph}(T)&=\fr{4\pi v}{gs}\int_0^\infty d\xi \Bigg\{s^2\Bigg[4\left(\fr{df}{d\xi}\right)^2+\fr{8}{\xi^2}f^2(1-f)^2+\fr{\xi^2}{2}\left(\fr{dh}{d\xi}\right)^2+h^2(1-f)^2\Bigg]\\
&+\fr{\xi^2}{g^2s^2}\Bigg[\fr{\la (T)}{4}s^4(h^4-1)-\fr{ET}{v} s^3(h^3-1)+\fr{D(T^2-T_0^2)}{v^2} s^2(h^2-1)\\
&+\fr{v^2}{8\La ^2}\Big[(s^6(h^6-1)-3s^4(h^4-1)+3s^2(h^2-1)\Big]\Bigg]\Bigg\}.
\end{aligned}
\]

The constant $C(T)$ is removed so that the sphaleron energy does not diverge. This is completely normal because this cosmological constant can be absorbed into the three normalized conditions of the effective potential. In order to ensure the convergence of sphaleron energy, we deliberately chose the ansatz $h(\xi)$ taking the form $\sim \xi^{-4}$ in the limit $\xi\rightarrow \infty$, so that after integrating we get the form $\sim \xi^{-1}$.

\begin{table}[h]
	\caption{Sphaleron energy at finite temperature using ansatz $a$, in units of $4\pi v/g \approx 4,738$ TeV with $\La =800$ GeV.}
	\centering
	\begin{tabular}{m{7em}m{6em}m{4em}m{4em}m{4em}c}
		\hline\hline
		$T$[GeV] & $v(T)$[GeV] & $E(T)$ & $a$ & $b$ & $E_{scaling}(T)$ (\% deviation)\\
		\hline
		$T_1=119.35$ & 138.77 & 1.035 & 2.853 & 3.612 & 1.122 (8.41\%)\\
		\hline
		119 & 144.47 & 1.087 & 2.785 &  3.411 & 1.168 (7.45\%)\\
		\hline
		$T_c=118.56$ & 149.78 & 1.136 & 2.738 & 3.272 & 1.211 (6.60\%)\\
		\hline
		117 & 162.32 & 1.249 & 2.646 & 3.008 & 1.312 (5.04\%)\\
		\hline
		116 & 168.02 & 1.301 & 2.617 & 2.927 & 1.359 (4.46\%)\\
		\hline
		115 & 172.78 & 1.344 & 2.591 & 2.855 & 1.397 (3.94\%)\\
		\hline
		$T'_0=113.36$ & 179.26 & 1.404 & 2.563 & 2.779 & 1.449 (3.21\%)\\
		\hline\hline
	\end{tabular}\label{bangspha1}
\end{table}

The result is given in the table \ref{bangspha1}. We see that the sphaleron energy decreases with increasing temperature. This is consistent with our expectation of the large baryon violation rate in the early universe. For comparison purposes, we showed in the last column the sphaleron energy calculated from the well-known scaling expression \cite{47, Ahriche1, Ahriche2}:
\be
	E_{scaling}(T)\approx \fr{v(T)}{v}E_{sph}(T=0).
\label{mar8}
\ee
In \cite{11}, the authors only used the above approximate expression and did not calculate directly the sphaleron energy at nonzero temperature from the effective potential.

\begin{figure}[!h]
		\centering
		\includegraphics[scale=0.7]{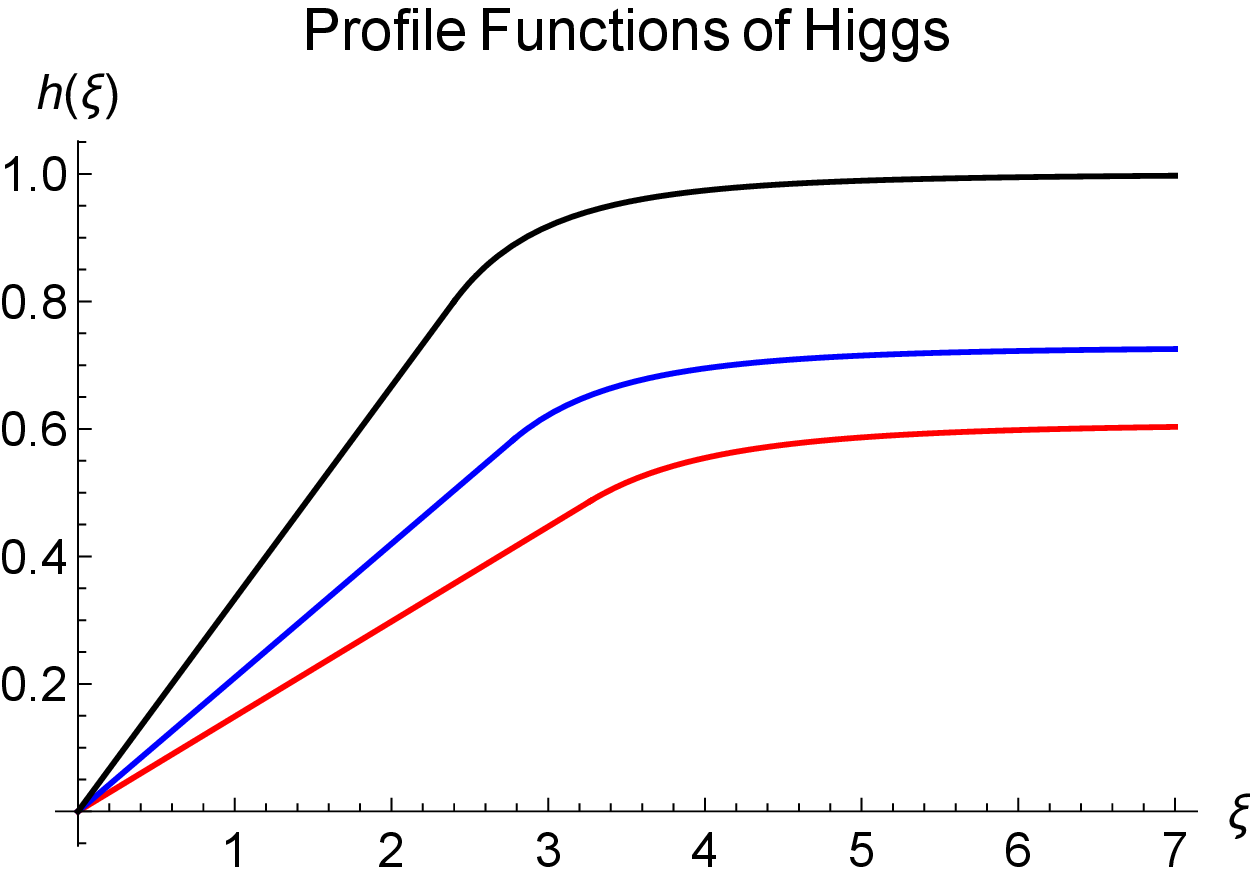}
		\caption{The radial functions of sphaleron for the Higgs field using ansatz $a$ with $\La =800$ GeV. Red line: $T=T_c$. Blue line: $T=T'_0$. Black line: $T=0$.}
		\label{sphaleron_h}
	\end{figure}

\begin{figure}[!h]
	\centering
	\includegraphics[scale=0.7]{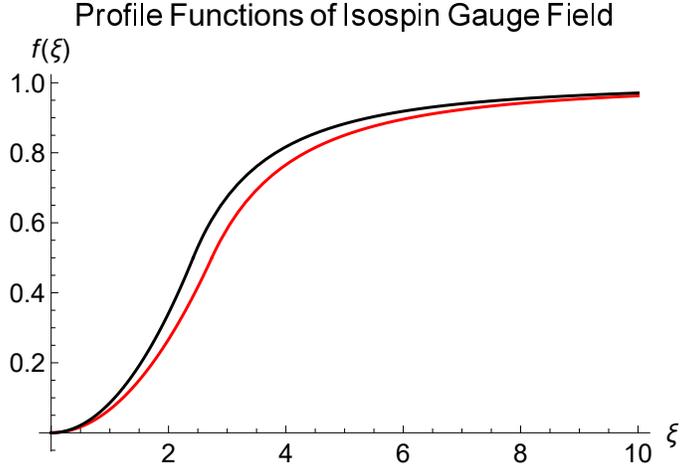}
	\caption{The radial function of sphaleron for the isospin gauge field using ansatz $a$ with $\La =800$ GeV. Red line: $T=T_c$. Black line: $T=0$.}
	\label{sphaleron_f}
\end{figure}

The profile functions of sphaleron for the Higgs boson and isospin gauge fields are plotted in the figures \ref{sphaleron_h} and \ref{sphaleron_f}, respectively.
\bc 
\ec 

The sphaleron solutions are only physically meaningful at $T\leq T_c$, although these solutions can be constructed mathematically up to $T_1$. Unlike the bubble solution, sphaleron solutions still exist at zero temperature and hence baryon number violation process is absent. The question of whether or not we can recreate the conditions of the early universe, and hence detect a process that violates baryon number in particle accelerators, is very deep and remains unanswered.

In order to ensure that the baryon number is preserved during the expansion of the universe, we need the baryon washout avoidance condition, which is also known as the decoupling condition or the third condition of Sakharov. The baryon-number violation rate must be smaller than the Hubble rate during the phase transition process (i.e. when the temperature drops from $T_c$ to $T_0$). On the other hand, EWPT occurs after inflation and belongs to the radiation-dominated period so that the Hubble rate is $H^2_{rad}=\fr{4\pi^3g_*}{45 m_{pl}^2}T^4$. The authors in Ref.\cite{11} carefully calculated the baryon number violation rate and gave results from the condition $\Ga_{sph} \ll H_{rad}$ as:
\be\label{decoupling_equation}
\fr{E_{sph}(T)}{T}-7ln\left(\fr{v(T)}{T}\right)+ln\left(\fr{T}{100 GeV}\right)>(35.9-42.8).
\ee

We see that the triples $\{T, v(T), E(T)\} $ at any temperatures in the table \ref{bangspha1} satisfy this condition. So at the cut-off scale of 800 GeV, the baryon number violation is preserved during the phase transition process. In other words, the condition of thermal imbalance has been satisfied.

\begin{figure}[!ht]
	\centering
	\includegraphics[scale=0.7]{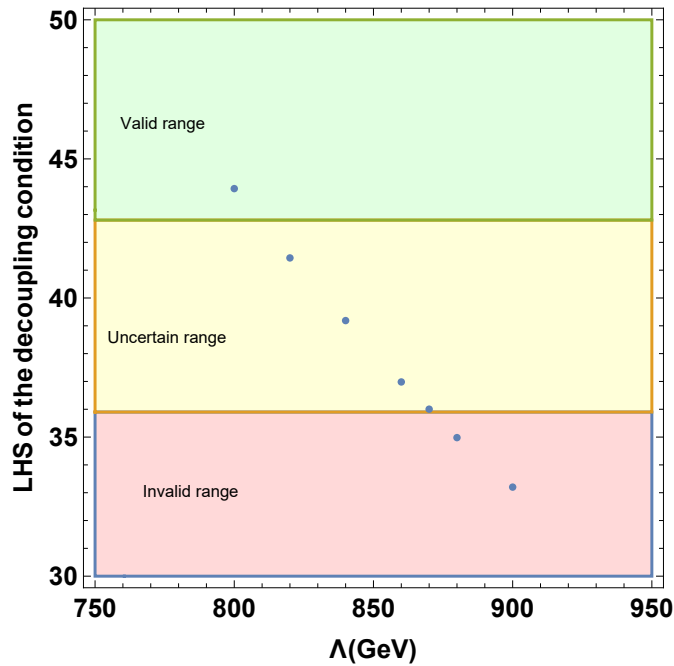}
	\caption{Description of the decoupling condition. Red zone: the region that does not satisfy the decoupling condition. Yellow zone: the error area corresponding to the right-hand side of Eq.\eq{decoupling_equation}. Green zone: the region that satisfies the decoupling condition.}
	\label{decoupling}
\end{figure}

Now we want to further evaluate the feasibility of the decoupling condition according to different cut-off values. We still use the ansatz $a$ and the sphaleron energy has been calculated at the critical temperature $T_c$ as in the table \ref{decoupling_table}. From the sets of values in the table \ref{decoupling_table}, the decoupling condition \eq{decoupling_equation} is visually depicted in the figure \ref{decoupling}.

\begin{table}[h!]
	\centering
	\caption{\{$T_c,v_c,E(T_c)$\} at different cut-off values.}
	\begin{tabular}{m{5em}m{5em}m{5em}c m{4em}m{4em}}
		\hline\hline
		$\La [GeV]$ & $T_c[GeV]$ & $v_c[GeV]$ & $E(T_c)\times\fr{4\pi v}{g}[TeV]$ & $a$ & $b$\\
		\hline
		900  & 126.68 & 114.22 & 0.862 & 2.765 & 3.350\\
		\hline
		880 &  125.24 & 120.79 & 0.912 & 2.764 & 3.349\\
		\hline
		870  & 124.48 & 124.38 & 0.940 & 2.763 & 3.346\\
		\hline
		860  & 123.71 & 127.81 & 0.966 & 2.762 & 3.342\\
		\hline
		840  & 122.08 & 135.21 & 1.023 & 2.754 & 3.319\\
		\hline
		820  & 120.38 & 142.36 & 1.078 & 2.748 & 3.301\\
		\hline
		800  & 118.56 &	149.78 & 1.136 & 2.738 & 3.272\\
		\hline\hline
	\end{tabular}
	\label{decoupling_table}
\end{table}

In summary, by calculating the sphaleron energy at finite temperature directly from the effective potential, we can give a more thorough assessment of the thermal imbalance condition, which is commonly known through $v_c/T_c >1$. This condition is only inherently reliable in the SM. In the previous section, we saw that the upper bound of the cut-off scale is 860 GeV to satisfy $v_c/T_c>1$. By assessing the  decoupling condition when calculating sphaleron energy we see that this upper bound is correct.

\section{Conclusion and discussion}\label{sec5}

With the same calculation as in the section \ref{sec4}, we can calculate the sphaleron energy with the contributions from other dimension-six operators as in Ref. \cite{11}. We also estimate that our calculation has an error not exceeding 10\% compared to the result calculated in Ref. \cite{11}.

These results are detailed and also consistent with the conclusions in the 2004 paper \cite{13} that explored the EWPT problem in the SM with the $\mathcal{O}_6 $ operator at 1-loop. In this article, the exact value of the Higgs mass is used to find the upper bound of $\La$ (about $860$ GeV) in order to satisfy the out-of-thermal equilibrium condition. It also shows the narrow temperature range where phase transition occurs.

Among many possible dimension-six operators, we only consider the $\mathcal{O} _6 $ operator for the Higgs potential. This is not a sufficient condition but a prerequisite for explaining matter-antimatter asymmetry  in the universe. Specifically, when the EWPT survey had $\mathcal {O} _6$, we found that the phase transition strength ($v_c/T_c$) is large enough to cause thermal nonequilibrium. In addition, when investigating the sphaleron solution, the baryon number violations are preserved during phase transition. Thus, the first and third conditions of Sakharov are feasible when considering $SM+ \mathcal{O}_6$. To fully solve the problem of matter-antimatter asymmetry, we need to cleverly combine the $\mathcal {O} _6$ operator with other dimension-six operators that cause C and CP violations. That is our possible future work.

We do not know if the contributions of these dimension-six operators are positive or negative, so we cannot conclude whether the sphaleron energy when calculating the addition of all dimension-six operators strongly satisfies the decoupling condition. However, in order to have a first-order phase transition, our calculation results show that the contribution from the dimension-six Higgs operator must be positive. This opens up the need to study Wilson's parameters more.

With the numerical result in \cite{11}, we find that in the allowed range of the cut-off scale \eq{mien} the contribution of $\mathcal{O}_6$ to the sphaleron energy at zero temperature does not exceed 5\% of the sphaleron energy of SM. If other dimension-six operators are also constrained in this domain and all of them have positive contributions, the total contribution will be about 35\%; this is a very significant contribution that will make the sphaleron energy increase greatly.

In the event that contributions of all dimension-six operators can compensate for each other, the sphaleron energy can only slightly increase compared to the SM. In a nutshell, there is a need for more information about the dimension-six operators by considering them in other quantum corrections or in the decay channels, which will allow us to calculate the sphaleron energy in more detail. In contrast, our range of the cut-off parameter and the EWPT and sphaleron problem are the good binding channels for finding new physics when adding dimension-six operators. Also, the dimension-eight operator $H^8$  has  been investigated   in Ref.\cite{chala}, while  the UV completions have been studied in Ref.  \cite{span,chala}. These are also our upcoming calculations as well as checking these solutions with the Cosmotransition  package \cite{cosmotran}.

Although numerical methods can be used to calculate the sphaleron energy without approximations for $J_{\mp}$, we cannot know the form of
 the effective potential. We do not know if it is minimal or not, and also $v(T)$ cannot be determined, so integrals can easily be diverged and give
  non-physical results. It is difficult to control these divergences by a numerical method. Therefore, if we use numerical methods without using
   approximations for $J$, we need a complex process to ensure the stability and physics of the solution.

In the calculation process of this paper, the "ansatz a" was carefully chosen to ensure the convergence of the sphaleron energy at finite
temperature at 1-loop level, providing that the effective potential takes the polynomial form. This form can only be obtained if we use the
 high-temperature approximation. If we did not use this approximation, the effective potential would have a very complicated form
 and therefore the convergence of the sphaleron energy would not be guaranteed anymore.

These results are a continuation of the previous articles, show a comprehensive view and more complete results for the cut-off parameter as well as the finite temperature electroweak sphaleron in the appropriate ansazts. It helps to show that the dimension-six operators can be one of the good improvements.

\section*{ACKNOWLEDGMENTS}
This research is funded by Vietnam  National Foundation for Science and Technology Development (NAFOSTED) under grant number 103.01-2019.346.
\\[0.3cm]

\end{document}